# Magnetoelectric coupling in $\varepsilon$-$Fe_2O_3$ nanoparticles

M. Gich*, C. Frontera, A. Roig, J. Fontcuberta, E. Molins

*Institut de Ciència de Materials de Barcelona-Consejo Superior de Investigaciones Científicas. Campus UAB, 08193 Bellaterra, Catalunya. Spain*

N. Bellido, Ch. Simon

*Laboratoire CRISMAT/ENSI-Caen, UMR 6508 du CNRS,6 Bd Marechal Juin, 14050 Caen, France.*

C. Fleta

*Institut de Microelectrónica de Barcelona (IMB-CNM, CSIC). Campus UAB, 08193 Bellaterra, Catalunya. Spain*

**Abstract**

Nanoparticles of the ferrimagnetic $\varepsilon$-$Fe_2O_3$ oxide have been synthesized by sol-gel method. Here, we report on the measurements of the dielectric permittivity as a function of temperature, frequency and magnetic field. It is found that, coinciding with the transition from collinear ferrimagnetic ordering to an incommensurate magnetic state occurring at about 100 K, there is an abrupt change (about 30 %) of permittivity suggesting the existence of a magnetoelectric coupling in this material. Indeed, magnetic field dependent measurements at 100 K have revealed an increase of the permittivity by about 0.3 % in 6 T. Prospective advantages of $\varepsilon$-$Fe_2O_3$ as multiferroic material are discussed.

* Email: mgich@icmab.es

Search for materials displaying coupled electric and magnetic properties, so called magnetoelectric materials, is nowadays driving a strong activity. Of particular interest are materials where ferromagnetic and ferroelectric behaviors coexist. Among them, the perovskites oxides such as $BiFeO_3$ [1] and $BiMnO_3$ [2] have received most attention. However, synthesis and crystallochemistry of Bi and Mn based perovskites is complex and stabilization of pure phase, accurate control of its composition and oxygen stoichiometry is far from undisputable [3]. Therefore, it appears that searching for alternative materials is a must and structures containing single-valent ions would be highly desirable. Among the iron (III) oxides, $\varepsilon$-$Fe_2O_3$ is a rare metastable polymorph [4, 5] which at room temperature is ferrimagnetic with a Curie temperature, $T_C$, of $T_C$ ~510 K and it presents an orthorhombic non-centrosymmetric structure (space group *Pna2₁*) [6]. So far, $\varepsilon$-$Fe_2O_3$ has only been stabilized in the form of nanoparticles and it is difficult to synthesize as a single phase, and thus, many of its properties remain unexplored. However, recent studies have disclosed a complex scenario for the $\varepsilon$-$Fe_2O_3$ magnetic properties. For instance, it has been found to exhibit a huge room-temperature coercivity of 20 kOe [7, 8]. In the context of the present paper, of more relevance is the existence of a transition, at about 100 K, to an incommensurate magnetic structure which is accompanied by a magnetic softening of the material [9]. Remarkably enough, the point group symmetry (*mm2*) of room-temperature $\varepsilon$-$Fe_2O_3$ structure is one of the so-called pyroelectric point groups thus implying the existence of permanent electric dipole moments and allowing the occurrence of physical effects such as optical activity, pyroelectricity, piezoelectricity or second harmonic generation [10]. Thus, the prospective coexistence in $\varepsilon$-$Fe_2O_3$ of both spontaneous magnetization and polarization makes the material attractive for magnetodielectric studies: an eventual coupling of these properties could find a widespread range of applications such as electric (magnetic) field tunable magnetic (dielectric) properties or multiple state memory elements. Since the discovery of magnetoelectric coupling in antiferromagnetic $Cr_2O_3$ [11], evidences of such effects were early reported on ferrimagnetic $GaFeO_3$ [12] which is isomorphous to $\varepsilon$-$Fe_2O_3$.

Here, we will report on the observation of an abrupt change of the dielectric permittivity in $\varepsilon$-$Fe_2O_3$ occuring simultaneously with the magnetic transition at $T$~100 K, thus suggesting the existence of a magnetoelectric coupling in the material. Subsequent measurements of the magnetic field dependence of the dielectric permittivity at 100 K have revealed relative changes of 0.3 % in 6 T. To the best of our knowledge this is the first report on the observation of magnetoelectric coupling in a pure ferric oxide.

The sample was prepared from iron nitrate (Riedel-de Haen, 96%) and tetraethoxysilane (TEOS) (Fluka, 98%) following a procedure similar to that described by Savii *et al.* [13]. Hydrolysis and condensation processes occurred in acidic hydroethanolic medium (TEOS:$H_2O$:EtOH = 1:6:6 mole ratio, reaction pH ~0.9). Gelation took place after 20 days at room temperature. The wet gel was dried at 60-80 ºC for 14 h, crushed and subsequently annealed three hours in air atmosphere, every 100 ºC up to 1100 ºC. The X-ray diffraction data revealed that after this preparation step, the material is an $\varepsilon$-$Fe_2O_3$/amorphous-$SiO_2$ composite where the iron oxide particles represent ~29 wt. % [9] and have a mean crystallite size of ~20 nm. In order to perform dielectric and magnetic measurements, the $SiO_2$ amorphous matrix was removed. The samples were treated in a concentrated NaOH solution at 80 ºC. After two days the solution was centrifuged and washed several times with distilled water until achieving a neutral pH and finally dried at 60 ºC. X-ray diffraction analysis performed after this treatment revealed the disappearance of the $SiO_2$ amorphous matrix while the $\varepsilon$-$Fe_2O_3$ phase remained unaltered.

Transmission electron microscopy (TEM) observations were carried out using a Phillips CM30 microscope operating at 300 kV. Neutron powder diffraction patterns were collected in the 4-488 K temperature range at ILL facility in Grenoble, using D1B beamline ($\lambda$=2.52 Å). The patterns were indexed using the *FullProf* [14] program. Magnetic measurements were performed using a Quantum Design SQUID magnetometer. For the dielectric measurements, the particles were pressed and shaped into a pellet of about 0.44 $cm^2$ and ~0.4 mm thick with a bulk density of 2.5 $g/cm^3$ which is about 50 % of the $\varepsilon$-$Fe_2O_3$ nominal density. The pellet was subsequently coated with silver paint to make a parallel plate capacitor. Dielectric measurements were performed using an Agilent4284A impedance analyzer by placing the sample inside a PPMS-Quantum Design cryostat to allow temperature and magnetic field control.

In Figure 1, the TEM images of both sample before and after the removal of the $SiO_2$ matrix show a distribution of roughly spherical isolated nanoparticles that are single crystals, well crystallized and have an average diameter ~ 20 nm. The structural characterization of the sample by means of X-ray diffraction, reported elsewere [9], revealed the only presence of $\varepsilon$-$Fe_2O_3$ nanoparticles with no traces of other ferric oxides polymorphs.

We have recently reported [9] a magnetic softening of $\varepsilon$-$Fe_2O_3$ at low-temperature (~90-120 K) that has been related to a transition from a collinear ferrimagnet to an incommensurate magnetic structure having a much reduced coercivity and a smaller magnetization. Figure 2a shows powder neutron diffraction patterns taken in the 80-160 K range. Exhaustive data analysis will be presented elsewhere[15]. For the purposes of this paper, the 0.5-2.5 $Å^{-1}$ *Q*-region is of major interest. It can be appreciated that on cooling, the intensity of the (011) reflection ($Q$ = 0.965 Å$^{-1}$) gradually lowers and simultaneously, the appearance of two satellites can be clearly observed. A similar trend can be identified for the weaker (120) ($Q$ = 1.873 Å$^{-1}$) reflection. This evolution indicates the development of an incommensurate magnetic structure; indexing of these satellites is achieved with a ~10 unit cell periodicity along the *b* axis [15]. In Fig. 2b we collect the temperature dependence of the integrated intensity of the (011) reflection and its satellites. Data in Fig. 2b indicates that the transformation from the collinear spin structure to the incommensurate one occurs in a quite broad temperature range ~ 80-110 K.

Lorenz *et al.* have recently demonstrated the existence of a magnetoelectric coupling in the orthorhombic phases of HoMnO$_3$ and YMnO$_3$ perovskites from anomalies in the dielectric and magnetic properties at the onset of an incommensurable magnetic order [16]. Similarly, in the hexagonal phases of these manganites, dielectric anomalies are detected at the onset of (frustrated) antiferromagnetic ordering [17]. Our observations suggest that a similar behavior could also be encountered in the ε-Fe$_2$O$_3$ system.

Indeed, we have found abrupt changes of the dielectric properties of ε-Fe$_2$O$_3$ at temperatures around the magnetic transition. In Fig. 3 we show the temperature dependence of the magnetization under 100 Oe of applied magnetic field (after a zero field cooling process); it is observed that upon cooling from the high temperature collinear ferrimagnetic state, a sudden decrease of magnetization occurs, starting at about 110 K and ending at about 85 K. No appreciable thermal hystheresis of the magnetization has been observed. We notice that the change of magnetization occurs at the same temperature range where the incommensurable magnetic order sets-in (see Fig. 2b). In Figure 3 we also include the temperature dependence of the dielectric constant obtained from impedance measurements. The raw capacitance measurements (2.45 nF at 40 K) allow us to estimate a relative permittivity of about 4 for $\varepsilon$-$Fe_2O_3$ at 40 K. More important here is the temperature dependence of the permittivity. To emphasize these changes, we plot in Fig. 3, $\Delta\varepsilon/\varepsilon_{min} = [\varepsilon(T)-\varepsilon_{min}]/\varepsilon_{min}$ where $\varepsilon_{min}$ is the minimal value of permittivity measured in the depicted temperature range. Data in Fig. 3 immediately reveal an abrupt change of $\Delta\varepsilon/\varepsilon_{min} \sim 30$ % occurring at the same temperature region where the magnetic transition sets in. The measured change of permittivity $\Delta\varepsilon/\varepsilon_{min} \sim 30$ % -between 80 and 140 K- is substantially larger than reported values for hexagonal rare earth manganites (~5 % at the Néel temperature $T_N$ of $YMnO_3$ for instance [17]) and comparable with that reported for the orthorhombic $YMnO_3$ (~ 60 % at the transition to the incommensurate antiferromagnetic state [16]).

From the microscopic point of view, the dielectric constant is related to both electronic and phononic excitations. The present experiments do not allow distinguishing between these two possible sources of dielectric changes and thus further experiments would be required. However, we may tentatively suggest that the weak spin-orbit coupling characteristic of $Fe^{3+}(3d^5)$ ions shall not promote substantial magnetoelastic effects and thus we do not expect significant structural changes at the transition from ferrimagnetic to incommensurable magnetic states. This suggestion is also supported by the absence, across the transition, of measurable position-shifts of the nuclear peaks of the neutron diffraction patterns (see for instance the (122) reflection in Fig. 2a) and the absence of thermal hysteresis in the magnetization measurements (Fig. 3). On the other hand, the substantial line-broadening that we have observed at about 100 K in preliminary Mössbauer spectroscopy measurements [15] is fully compatible with an electronic density redistribution. Data in Figs. 3 and 4a show that the permittivity is reduced in the incommensurate low-temperature magnetic phase. The dielectric permittivity depends on the initial and final states of the electronic configuration as $\varepsilon \sim |\langle i|E|f\rangle|^2/V_o^2$, where $|i\rangle$, $|f\rangle$ and $V_o$ represent the initial and final states and the bandgap respectively and E is the applied electric field. We note that electronic excitations in a magnetic material are spin-dependent and thus the matrix element $\langle i|E|f\rangle$ and the energy gap shall depend on the spin texture. Consequently, it is not trivial to predict the sign of the $\varepsilon$ variation through a magnetic transition. However, these considerations provide a microscopic reason for the dependence of permittivity on the magnetic state. A similar scenario was proposed for $RMnO_3$ [17] where, interestingly enough, it was also found that $\varepsilon$ lowers at the Néel temperature. Therefore, we suggest that spin-dependent electronic excitations may dominate the changes of permittivity at around 100 K.

Impedance measurements were also performed at several frequencies between 5 kHz and 100 kHz and it has been found that the absolute change of permittivity $\Delta\varepsilon/\varepsilon_{min}$ is weakly dependent on frequency. In Figure 4a, we present the measurements at some representative frequencies of the temperature dependence of the capacitance $C/C_0$ (main panel) and conductance $G/G_0$ (lower inset) relative to the value obtained at 80 K. We note that upon increasing frequency, the temperatures at which the $\Delta\varepsilon(T)$ jump occurs and where the relative conductance displays a maximum, are both shifted to higher temperature. This behavior is expected in a system where dielectric polarization relaxation follows a Debye model, and where the characteristic relaxation time has a thermally activated dependence [18]. In the upper inset of Fig. 4a, we plot the dependence of $T_{max}$, the temperature at which the maximum of $G/G_0$ occurs, versus the measuring frequency, f. It is clear that in the narrow temperature range around the magnetic transition, the data follow an Arrhenius-like form, evidencing that an activated relaxation process may account for the observed dependence of the dielectric response on frequency. From the slope of data in Fig. 4a (upper inset) one can extract an activation energy $E_{act}$≈0.13 eV and a pre-exponential time factor $\tau_o$≈ 6.4·10$^{-11}$ s. It is interesting to note that the activation energy $E_{act}$≈0.13 eV is much larger than the magnetic energy involved in a transition occurring at about 100 K (~8 meV) but it is a substantial fraction of the charge transfer gap in an insulating ferric oxide. This observation is in agreement with the suggestion of a dominant role of electronic excitations in the observed changes of permittivity.

One should also note in Fig. 4a, that at temperatures far above or below the transition, the frequency dependence of $C/C_0$ is almost washed out, thus suggesting that the relevant energy scale is temperature dependent -and thus the Arrhenius behavior holds only in limited temperature range- and becomes exceedingly large away from the magnetic transition. In fact, this behavior could be anticipated based on quite general grounds; indeed, if $U_{HT}$ and $U_{LT}$ correspond to the free energies of the high-temperature (HT) and low-temperature (LT) magnetic phases, and assuming a linear temperature dependence, then $U_{HT} = U_{HT}^0(1+B_{HT}T)$ and $U_{LT} = U_{LT}^0 (1+B_{LT}T)$ where $U^0$ and B are appropriate coefficients; therefore, the relevant activation energy can be estimated as $\Delta U = U_{HT} - U_{LT}$ which will obviously be temperature dependent. Only in the temperature region around 100 K, $\Delta U$ is of the order of the excitation (or the relaxation time is within the experimental window) and it is only there that we observed a frequency-dependent response.

Data in Fig. 3 and 4a indicate that the dielectric permittivity of ε-Fe$_2$O$_3$ is extremely sensitive to magnetic order and thus magnetocapacitance effects are to be expected. In order to get a direct evidence of this coupling we have measured the magnetic-field dependence of the capacitance at various temperatures. In Fig. 4b we show the magnetoelectric response ($\Delta C/C(0)=C(H)-C(0)/C(0)$) of ε-Fe$_2$O$_3$ measured at 100K (at f = 100 kHz). Data in Fig. 4b shows a change of capacitance of about 0.3% at 6T. It is appropriate to recall that similar magnetocapacitance changes have been found in polycrystalline orthorhombic YMnO$_3$ [16] and BiMnO$_3$ [19].

It is interesting to note that the complete capacitance cycle measured up to 6 T does not show significant hysteresis and mimics the magnetization loop measured at the same temperature (100 K) where the coercivity of ε-Fe$_2$O$_3$ vanishes [9]. Interestingly enough, measurements performed at somewhat higher temperatures (not shown) indicate hysteretic behavior of both magnetization [9] and capacitance, thus confirming the occurrence of magnetoelectric coupling. Following Smolenskii [20], Kimura et al. [19] showed that, starting from a phenomenological Ginzburg-Landau model in presence of magnetoelectric coupling, at temperatures close to the magnetic transition the relative change of dielectric permittivity should be proportional to the square of the magnetization: $\delta\varepsilon \sim \gamma M^2$, where $\gamma$ is the magnetoelectric coupling coefficient. In order to test this prediction, we have plotted the measured relative change of capacitance $\Delta C/C(0) \sim \Delta\varepsilon/\varepsilon(0)$ *vs* $M^2$, at some temperatures around the transition (100 K and 110 K) (Fig. 4b-inset). As shown, the data displays a rough linear behavior, thus implying that $\Delta C/C_O = \gamma^* M^2$. At temperatures far away from 100 K, the linear behavior no longer holds, as found in BiMnO$_3$ [20]. This may indicate [20] that the magnetic hardening [9] and the corresponding strong magnetic pinning affect the dielectric constant. Therefore, data in Fig. 4b indicate the presence of a magnetoelectric coupling and the Fig. 4b-inset suggests that this coupling can be described in terms of the Ginzburg-Landau model.

In conclusion, we have shown that magnetic and dielectric properties of pure ε-Fe$_2$O$_3$ nanoparticles are coupled. To our knowledge, this is the first time that such kind of behavior is reported for a single-metal ferrimagnetic oxide. A ferric oxide such as ε-Fe$_2$O$_3$ would have significant advantages over the Bi or Mn-based biferroic materials, in terms of control of stroichiometry and current leakages due to its more stable chemical composition as well as from an economical point of view. Although preparation of ε-Fe$_2$O$_3$ oxide in functional form such as epitaxial films may be challenging, the results presented here shall stimulate new work and open the door to future investigations which can be promising for applications.

**Acknowledgements:** Financial support from Ministerio de Educación y Ciencia (Spain), Project MAT2003-01052 and Generalitat de Catalunya, Project 2001SGR00335 are gratefully acknowledged. Neutron diffraction experiments were performed under the CRG-D1B España programme of the Ministerio de Educación y Ciencia (MEC). C. F. acknowledges financial support from MEC (Spain). We are greatly indebted to M. Lozano (CNM-CSIC) for their assistance during impedance measurements, C. Bonafos and E. Snoeck (CEMES-CNRS) who performed most TEM measurements and J. Casabó and Ll. Escriche (Departament de Química, Universitat Autònoma de Barcelona) for their assistance in the removal of the SiO$_2$ amorphous matrix.

**FIGURE CAPTIONS**

**FIG. 1:** TEM images of the sample before and after removing the $SiO_2$ matrix (top). The HRTEM image (bottom) shows a defect-free, single-crystalline $\varepsilon$-$Fe_2O_3$ nanoparticle.

**FIG. 2: (a)** Powder neutron diffraction patterns of $\varepsilon$-$Fe_2O_3$ in the 80-160 K range. **(b)** Temperature dependence of the integrated intensity of the (011) reflection and its satellites as deduced from the profile refinement.

**FIG. 3:** Dependence of the magnetization on temperature after cooling in zero field ($\triangledown$) and relative change of permittivity for $\varepsilon$-$Fe_2O_3$ (■)

**FIG. 4: (a)** Temperature dependence of the capacitance normalized to the capacitance value at 80 K ($C_O$) measured at different frequencies. The lower inset shows the corresponding conductance normalized to the $G_O$ values. The upper inset shows the frequency dependence for the temperature of the conductance maximum. **(b)** Magnetic field induced change of the capacitance $\Delta[C/C(0) = C(H)-C(0)/(C0)]$ at 100 K in a complete cycle; (■) and (□) indicate values recorded decreasing and increasing field respectively. Inset: magnetic field induced change of the capacitance as a function of the squared magnetization at temperatures around the magnetic transition: 100 K and 110 K (closed and open symbols respectively).

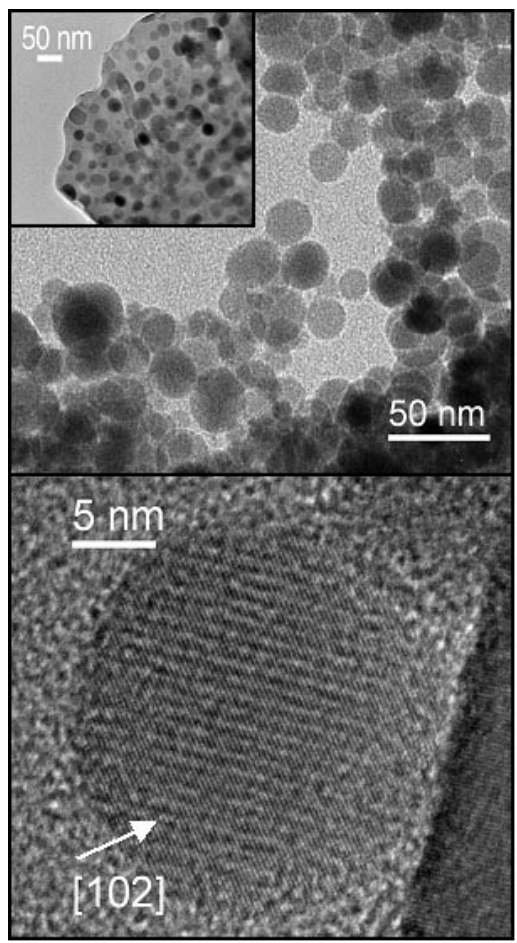

Figure 1 M. Gich et al.

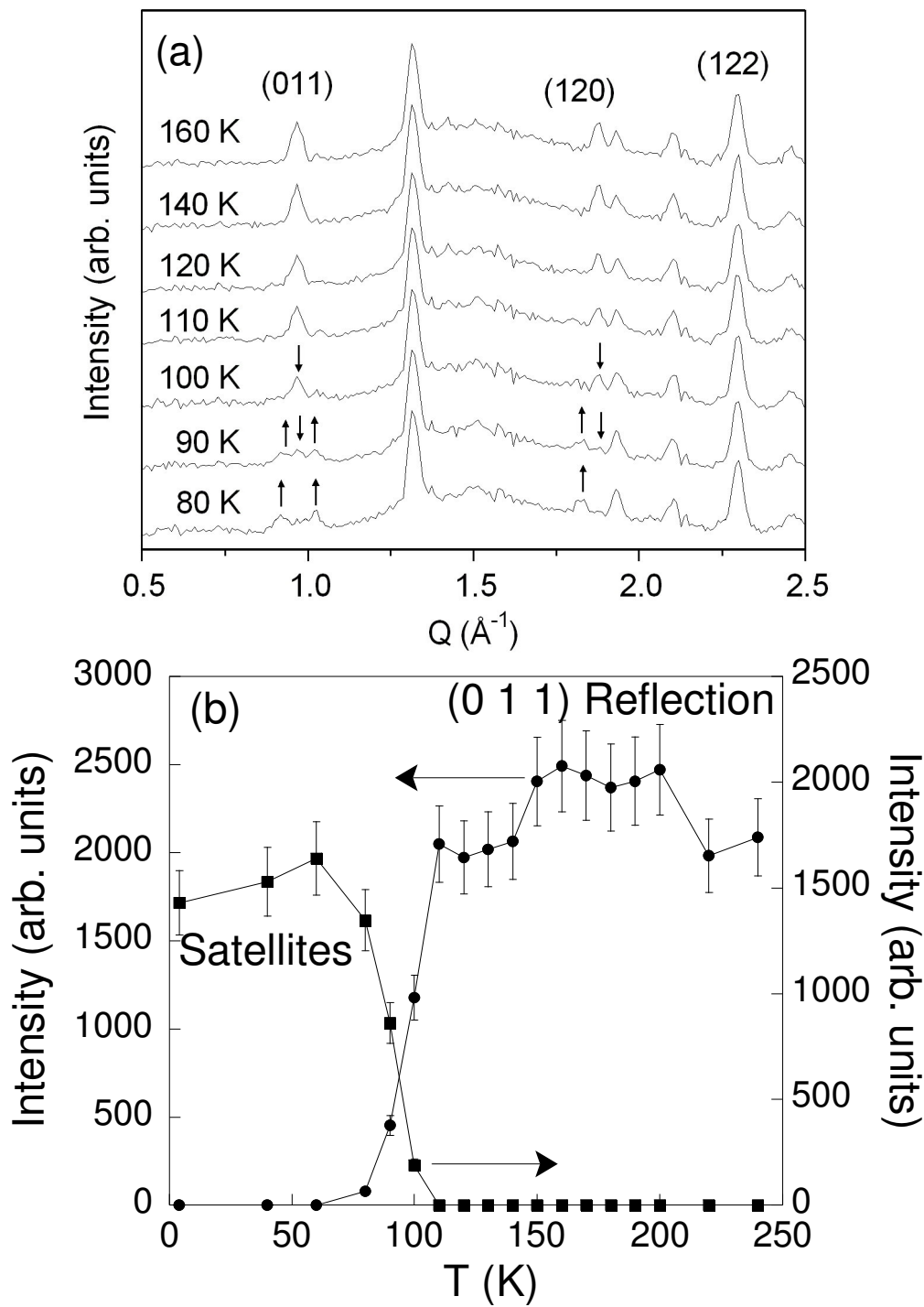

Figure 2　　　　　　　　　　　　　　　　　　　　　　　　M. Gich et al.

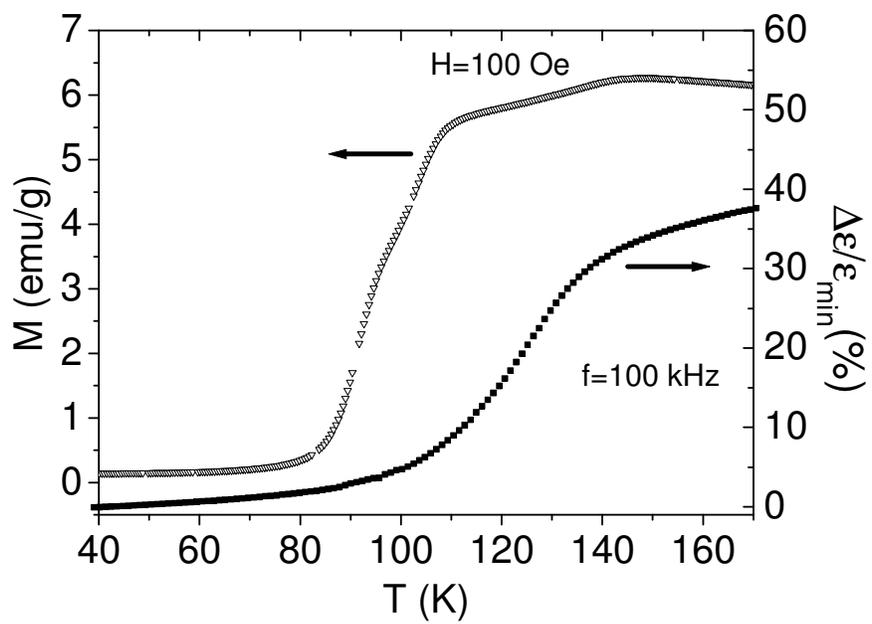

Figure 3 M. Gich et al.

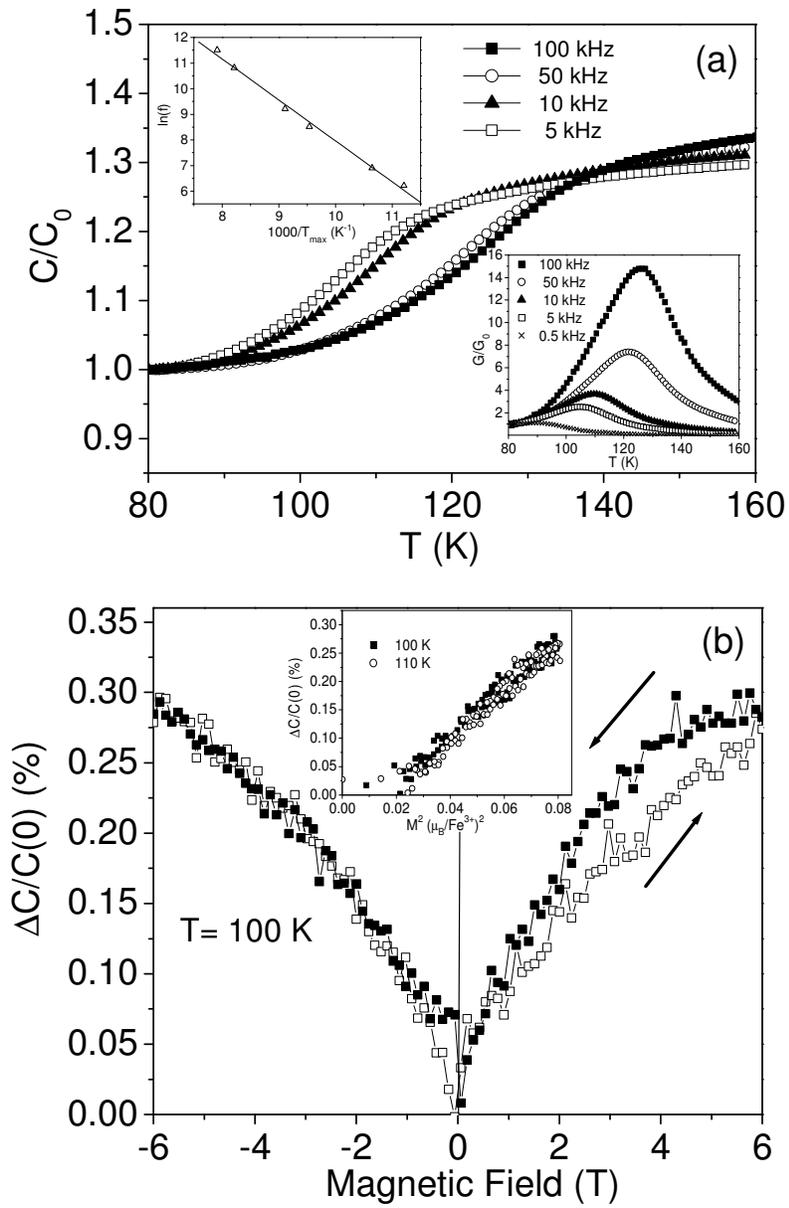

Figure 4 M. Gich et al.

17